\documentclass[12pt]{iopart}

\usepackage{cite}
\usepackage{graphicx}
\usepackage{multirow}

\usepackage{color}

\begin{document}

\title{Noncyclic and nonadiabatic geometric phase for counting statistics}

\author{Jun Ohkubo\footnote{
Present address: Department of Systems Science, Graduate School of Informatics, Kyoto University, 36-1,
Yoshida Hon-machi, Sakyo-ku, Kyoto-shi, Kyoto 606-8501, Japan.
} and Thomas Eggel}

\address{
Institute for Solid State Physics, University of Tokyo, 
5-1-5, Kashiwanoha, Kashiwa, Chiba 277-8581, Japan
}
\ead{ohkubo@i.kyoto-u.ac.jp}
\begin{abstract}
We propose a general framework of the geometric-phase interpretation for counting statistics.
Counting statistics is a scheme to count the number of specific transitions in a stochastic process.
The cumulant generating function for the counting statistics can be interpreted as a `phase',
and it is generally divided into two parts: the dynamical phase and a remaining one.
It has already been shown that for cyclic evolution the remaining phase corresponds to a geometric phase,
such as the Berry phase or Aharonov-Anandan phase.
We here show that the remaining phase also has an interpretation as a geometric phase
even in noncyclic and nonadiabatic evolution.
\end{abstract}

% 03.65.Vf Phases: geometric; dynamic or topological 
% 05.40.-a Fluctuation phenomena, random processes, noise, and Brownian motion
% 02.50 Ey Stochastic processes 
% (05.10.Gg Stochastic analysis methods (Fokker-Planck, Langevin, etc.) )
\maketitle

\section{Introduction}

Numerous investigations of nonequilibrium physics have been carried out recently,
and some general results have already been found.
For example, the fluctuation theorem 
\cite{Gallavotti1995,Gallavotti1995a,Kurchan1998,Crooks1998,Crooks1999,Lebowitz1999,
Maes1999,Crooks2000,Gaspard2004,Seifert2005}, 
the Jarzynski equality \cite{Jarzynski1997,Jarzynski1997a},
and the Hatano-Sasa relation \cite{Hatano2001} are examples of recent developments
in nonequilibrium physics.
These concepts are related to each other \cite{Esposito2007,Seifert2008,Ohkubo2009},
and then one could imagine that they may include interesting mathematical structures.
Moreover, if there are mathematically interesting structures,
it could be expected that these mathematical structures are also physically-meaningful.
Hence, it would be valuable to seek mathematical structures
behind nonequilibrium states.

We here focus on stochastic processes described by master equations,
and try to find mathematical structures behind them.
In general, a master equation describes the time-development of the probability of states.
One of the important quantities for nonequilibrium states is the `current',
which is related to the number of specific transitions occurring during a given time.
For example, in a particle hopping model on a one dimensional lattice,
the numbers of particles moving in and out of a specific site gives information about the current.
In addition, in order to evaluate all statistics, including fluctuations, of the number of 
specific transitions,
the framework of counting statistics is available 
\cite{Bicout1999,Gopich2003,Gopich2005,Sung2005,Gopich2006}.
Actually, the counting statistics for stochastic processes
enables one to obtain information about nonequilibrium currents 
in a simple two-state model under cyclic perturbations,
for both adiabatic regimes \cite{Sinitsyn2007,Sinitsyn2007a,Sinitsyn2008,Sinitsyn2009}
and nonadiabatic regimes \cite{Ohkubo2008,Ohkubo2008a};
here, `adiabatic' means that the perturbations are performed very slowly.
In general, a cumulant generating function for the counting statistics
can be interpreted as a `phase', as discussed in \cite{Sinitsyn2007}.
The phase is divided into two parts, i.e., the dynamical phase and a remaining one.
In previous studies
\cite{Sinitsyn2007,Sinitsyn2007a,Sinitsyn2008,Sinitsyn2009,Ohkubo2008,Ohkubo2008a}, 
it has been shown that the concept of geometric phase,
such as the Berry phase and the Aharonov-Anandan phase, is useful to calculate
the current statistics due to cyclic perturbations.

While a \textit{noncyclic} geometric phase has already been studied in quantum mechanics 
\cite{Samuel1988,Aitchison1992,Mukanda1993,Mostafazadeh1999},
the geometric phase discussion in the counting statistics 
has been extended to \textit{noncyclic} cases recently \cite{Sinitsyn2008a,Sinitsyn2010}.
However, the discussions in \cite{Sinitsyn2008a,Sinitsyn2010} are 
restricted to adiabatic cases, which correspond to quasi-static evolutions.

The aim of the present paper is to develop 
a geometric discussion for noncyclic and \textit{nonadiabatic} cases.
In order to investigate transitions between two nonequilibrium states, 
one has to consider nonadiabatic and noncyclic evolutions.
In the present paper, we will show that the remaining phase has 
a geometric interpretation even in the nonadiabatic and noncyclic cases,
based on the discussions given by Samuel and Bhandari for quantum systems \cite{Samuel1988}.
Hence, the present paper will give a natural extension of the adiabatic case
given in \cite{Sinitsyn2008a,Sinitsyn2010} to the nonadiabatic case.

The present paper is constructed as follows.
In section 2 we briefly explain the general scheme for counting statistics,
and rewrite the formulation in terms of bra and ket state vectors.
Section 3 is a brief explanation for cyclic evolution cases.
Section 4 is the main part of the present paper;
we confirm that the remaining phase in noncyclic cases is adequately interpreted as a geometric phase.
In section 5, 
we give concluding remarks and some discussions about physical meanings of the noncyclic geometric phase.

\section{Time evolution equation for counting statistics}

\subsection{Basics for counting statistics}

We here briefly review counting statistics for stochastic processes.
The formulation is based on a usual generating function approach,
which has been widely used in various contexts
(e.g., for the charge transport or full counting statistics in quantum systems, see \cite{Weiss_book}.)
Details concerning the counting statistics can be found 
in the Appendix or, for example, in \cite{Gopich2005}.

Let us start from a master equation, which describes the time evolution of the probability distribution.
The master equation is given by
\begin{eqnarray}
\frac{\rmd}{\rmd t} p_n(t) = \sum_{m} K_{nm}(t) p_m(t),
\label{eq_master_equation}
\end{eqnarray}
where $p_n(t)$ is the probability of finding the system in state $n$,
and $K_{nm}(t)$ is the rate constant for transition $m \to n$.
In general, the transition matrix $\{K_{nm}(t)\}$ is time-dependent.

The aim of the counting statistics is to count the numbers of target transitions.
For simplicity, we here consider a case in which
the number of target transitions $i_\mathrm{A} \to j_\mathrm{A}$ is counted.
It is easy to extend the following discussions to multiple transition cases,
and in these cases it is possible to derive a joint probability distribution 
for several target transitions.

In order to extract statistics for the target transition $i_\mathrm{A} \to j_\mathrm{A}$,
a generating function is defined as
\begin{eqnarray}
F(\chi,t) \equiv \sum_{k=0}^\infty \rme^{k \chi} P(k|t),
\end{eqnarray}
where $P(k|t)$ is the probability of $k$ target transitions $i_\mathrm{A} \to j_\mathrm{A}$
during the time $t$.
Derivatives of $F(\chi,t)$ with respect to $\chi$ give
various information about the statistics of the target transition.
As shown in Appendix,
the generating function $F(\chi,t)$ is constructed 
from a linear combination of functions $\{f_n(\chi,t)\}$:
\begin{eqnarray}
F(\chi,t) = \sum_n f_n(\chi,t).
\label{eq_redef_of_total_generating_function}
\end{eqnarray}
$f_n(\chi,t)$ is a \textit{restricted} generating function under the condition that
the system is in state $n$ at time $t$,
and it satisfies the following time-evolution equation:
\begin{eqnarray}
\frac{\rmd}{\rmd t} f_n(\chi,t) = \sum_{m} \kappa_{nm}(\chi,t) f_m(\chi,t).
\label{eq_gf_original}
\end{eqnarray}
The matrix $\{\kappa_{nm}(\chi,t)\}$ is a \textit{modified} transition matrix
obtained by introducing counting parameters $\chi$
into the transition matrix $\{K_{nm}(t)\}$;
the off-diagonal element $K_{j_\mathrm{A} i_\mathrm{A}}(t)$,
which is the rate constant for the target transition,
is replaced by $K_{j_\mathrm{A} i_\mathrm{A}}(t) \rme^{\chi}$ \cite{note1}.
Note that the modified transition matrix becomes the same as the transition matrix if we set $\chi = 0$,
i.e., $\{\kappa_{nm}(\chi=0,t)\} = \{K_{nm}(t)\}$.
In addition, \eref{eq_gf_original} should be solved
using the initial condition $f_n(\chi,0) = p_n(0)$ (see Appendix).

\subsection{Expressions in terms of bra and ket state vectors}

The time-evolution equations for the restricted generating functions \eref{eq_gf_original}
can be rewritten in terms of a `ket' state vector:
\begin{eqnarray}
\frac{\partial}{\partial t} | \psi(\chi,t) \rangle
= H(\chi,t) | \psi(\chi,t) \rangle
\label{eq_time_evolution_for_ket},
\end{eqnarray}
where $H(\chi,t) \equiv \{\kappa_{nm}(\chi,t)\}$
is the modified transition matrix.
The ket state vector $|\psi(\chi,t)\rangle$ is defined as
$|\psi(\chi,t)\rangle \equiv (f_1(\chi,t), f_2(\chi,t), \dots )^\mathrm{T}$,
where the superscript $(\cdot)^\mathrm{T}$ denotes the transposed matrix.

Since all solutions to \eref{eq_time_evolution_for_ket} form a linear vector space,
there is a dual space.
In fact, we can construct the adjoint equation of \eref{eq_time_evolution_for_ket} as follows:
\begin{eqnarray}
\frac{\partial}{\partial t} \langle \psi(\chi,t) |
= - \langle \psi(\chi,t) | H(\chi,t).
\label{eq_time_evolution_for_bra}
\end{eqnarray}
Note that $\langle \psi(\chi,t)|$ is not the complex conjugate of $| \psi(\chi,t) \rangle$,
because of the non-Hermitian property of $H(\chi,t)$.
In addition, we can show that the norm, i.e, the inner product
between $\langle \psi(\chi,t)|$ and $| \psi(\chi,t)\rangle$,
is preserved:
\begin{eqnarray}
\frac{\rmd}{\rmd t} \langle \psi(\chi,t) | \psi(\chi,t) \rangle = 0.
\label{eq_conserved_norm_original}
\end{eqnarray}
This conserved norm is a consequence of the minus sign in the adjoint system 
\eref{eq_time_evolution_for_bra}.

When the bra vector at the initial time is 
$\langle \psi(\chi,0)| = (1, 1, 1, \dots)$,
the generating function $F(\chi,t)$ is expressed in terms of the bra and ket state vectors as follows:
\begin{eqnarray}
F(\chi,t) = \langle \psi(\chi,0) | \psi(\chi,t) \rangle.
\end{eqnarray}
This fact means that
it is sufficient to compare the bra state vector at the initial time, $\langle \psi(\chi,0)|$,
and the ket state vector at time $t$, $| \psi(\chi,t) \rangle$,
in order to obtain the generating function.
Additionally,
using the initial conditions for the bra and ket state vectors,
and recalling the conserved norm \eref{eq_conserved_norm_original},
we obtain
\begin{eqnarray}
\langle \psi(\chi,t) | \psi(\chi,t) \rangle = 
\langle \psi(\chi,0) | \psi(\chi,0) \rangle = 1.
\label{eq_conserved_norm_original_2}
\end{eqnarray}

\section{Cyclic evolution}

If the system develops in a cyclic way,
the discussions become easy.
For the cyclic case with period $T$,
the cyclic state is written as
\begin{eqnarray}
|\psi(\chi,T)\rangle = \rme^{\eta(\chi)} | \psi(\chi,0) \rangle,
\end{eqnarray}
i.e., the ket state vector at time $T$ is proportional to that at the initial time.
Note that $\eta(\chi)$ is a real number
due to the construction of the modified transition matrix $H(\chi,t)$.
We here call $\eta(\chi)$ a `phase'.
The reason  for this choice of terminology will be explained later.

The generating function for the cyclic case is simply expressed as
\begin{eqnarray}
F(\chi,T) = \langle \psi(\chi,0) | \rme^{\eta(\chi)} |\psi(\chi,0) \rangle
= \rme^{\eta(\chi)},
\end{eqnarray}
where we used the conserved norm \eref{eq_conserved_norm_original_2}.
Hence, the phase $\eta(\chi)$ is the cumulant generating function.
As shown in 
\cite{Sinitsyn2007,Sinitsyn2007a,Sinitsyn2008,Sinitsyn2009,Ohkubo2008,Ohkubo2008a},
the cumulant generating function is divided into two parts,
the dynamical part and a geometric part.
The geometric part can be interpreted as the Berry phase or Aharonov-Anandan phase
for the adiabatic or nonadiabatic evolutions, respectively.
It has also been clarified that
these geometric phases are physically-meaningful,
and they are deeply related to the pumping phenomena discussed in 
\cite{Sinitsyn2007,Sinitsyn2007a,Sinitsyn2008,Sinitsyn2009,Ohkubo2008,Ohkubo2008a}.

We here comment on the reason of calling the quantity $\eta(\chi)$ a `phase'.
If one considers a Schr{\"o}dinger equation with Hermitian Hamiltonian,
a phase $\zeta$ is defined as $\exp(\rmi \zeta)$, where $\zeta$ is a real number.
However, in general cases with non-Hermitian Hamiltonian, 
the phase becomes a \textit{complex} phase \cite{Garrison1988}.
In this sense, the phase $\eta(\chi)$ is considered as a special case of a complex phase.
In addition, the dynamical phase and geometric phase 
are adequately defined 
in a similar way to that in quantum mechanics.
Hence, the term `phase' is used.

\section{Geometric phase for noncyclic and nonadiabatic evolution}

For noncyclic cases,
we cannot use the geometric phase concept developed in 
\cite{Sinitsyn2007,Sinitsyn2007a,Sinitsyn2008,Sinitsyn2009,Ohkubo2008,Ohkubo2008a}.
The difficulty basically stems from the fact that
the state vector at time $t$ is \textit{not} proportional
to that at the initial time.
Hence, it is impossible to simply compare the `phase' difference
between the two states.

We here develop a general framework of a nonadiabatic and noncyclic geometric phase for counting statistics.
After giving some discussions about dynamical and remaining phases,
we will show the reason why the remaining phase can be interpreted as
a \textit{geometric} phase,
following closely the discussion given by Samuel and Bhandari \cite{Samuel1988}.

\subsection{Definition of the remaining phase}

First, we define the dynamical phase as usual:
\begin{eqnarray}
\delta(\chi,t') \equiv \langle \psi(\chi,t') | H(\chi,t') | \psi(\chi,t') \rangle.
\end{eqnarray}
The new state vectors, from which the dynamical phase is removed,
are defined as follows:
\begin{eqnarray}
| \phi(\chi,t) \rangle \equiv \exp\left( - \int_0^t \delta(\chi,t') \rmd t' \right) | \psi(\chi,t) \rangle, \\
\langle \varphi(\chi,t) | \equiv \exp\left( \int_0^t \delta(\chi,t') \rmd t' \right) \langle \psi(\chi,t) |. 
\end{eqnarray}
As discussed in section 2.2,
the norm is conserved and hence we obtain
\begin{eqnarray}
\frac{\rmd}{\rmd t} \langle \varphi(\chi,t) | \phi(\chi,t) \rangle = 0,
\qquad  \langle \varphi(\chi,t) | \phi(\chi,t) \rangle = 1.
\end{eqnarray}
Note that $\delta(\chi,0) = 0$.

Using the new state vectors,
the generating function can be rewritten as
\begin{eqnarray}
F(\chi,t)
&= \langle \varphi(\chi,0) | \exp\left[\int_0^t \delta(\chi,t') \rmd t' \right] | \phi(\chi,t) \rangle 
\nonumber \\
&= \exp\left[\int_0^t \delta(\chi,t') \rmd t' \right]
 \exp [ \ln \langle \varphi(\chi,0) | \phi(\chi,t) \rangle ].
\end{eqnarray}
The total phase $\mu(\chi,t)$, defined as $\rme^{\mu(\chi,t)} \equiv F(\chi,t)$,
is adequately divided into two parts:
\begin{eqnarray}
\mu(\chi,t) = \int_0^t \delta(\chi,t') \rmd t' + 
\ln \langle \varphi(\chi,0) | \phi(\chi,t) \rangle.
\end{eqnarray}
Hence, the remaining phase $\gamma(\chi,t)$ is defined as 
\begin{eqnarray}
\gamma(\chi,t) \equiv \ln \langle \varphi(\chi,0) | \phi(\chi,t) \rangle.
\label{eq_def_of_gamma}
\end{eqnarray}
If the solutions of \eref{eq_time_evolution_for_ket} and \eref{eq_time_evolution_for_bra} 
are obtained,
it is possible to estimate these two phases.

Although the two `phases' were defined,
one may wonder whether this division is important and meaningful or not.
As is easily seen, these phases correspond directly to the cumulant generating functions.
Hence, one may expect that
the remaining phase is also meaningful,
in a similar way to the cyclic cases in section 3.
In the cyclic cases, the remaining phase can be interpreted as
a geometric phase, and it gives important information about the pump current
\cite{Sinitsyn2007,Sinitsyn2007a,Sinitsyn2008,Sinitsyn2009,Ohkubo2008,Ohkubo2008a}.
In the next subsection, we will show that 
it is actually possible to interpret the phase $\gamma(\chi,t)$ 
as a noncyclic \textit{geometric} phase.

\subsection{Interpretation of the remaining phase as a noncyclic geometric phase}

Let $\mathcal{N}$ denote the set of all states $| \phi(\chi,t)\rangle$:
$\mathcal{N} = \{ | \phi(\chi,t) \rangle \}$.
$\mathcal{R}$ is the space of rays: $\mathcal{R} = \mathcal{N} / \sim$,
where $\sim$ denotes that elements of $\mathcal{N}$, which differ only by a phase,
are regarded as equivalent.
In the present paper, the phase is always defined as a real number 
because of the construction of the modified transition matrix.
Hence, the relation $\sim$ means that an element of $\mathcal{N}$ 
is proportional to another element of $\mathcal{N}$.
A natural projection map $\pi: \mathcal{N} \to \mathcal{R}$ connects each vector
to the ray on which it lies.
The triplet $(\mathcal{N},\mathcal{R},\pi)$ forms a principal fiber bundle
over the base space $\mathcal{R}$.
Since $|\phi(\chi,t)\rangle$ satisfies 
\begin{eqnarray}
\frac{\rmd}{\rmd t} | \phi(\chi,t) \rangle
= \left[ H(\chi,t) - \delta(\chi,t) \right] | \phi(\chi,t) \rangle,
\label{eq_time_evolution_for_phi}
\end{eqnarray}
we obtain the following parallel transport law
\begin{eqnarray}
\langle \varphi(\chi,t) | \frac{\rmd}{\rmd t} | \phi(\chi,t) \rangle = 0,
\end{eqnarray}
which defines a natural connection on this fiber bundle.

Let $| \phi(\chi,s) \rangle$ be a curve in $\mathcal{N}$.
The tangent vector to this curve is defined as $| u \rangle = (\rmd/\rmd s) | \phi (\chi,s) \rangle$.
The corresponding tangent vector in the dual space is 
$\langle v | = (\rmd/\rmd s) \langle \varphi (\chi,s) |$.
We here introduce a quantity
\begin{eqnarray}
A(\chi,s) = \langle \varphi(\chi,s) | u (\chi,s) \rangle.
\end{eqnarray}
For a \textit{cyclic} case,
the Aharonov-Anandan phase is defined by using $A(\chi,s)$.
In addition, 
introducing gauge transformations,
\begin{eqnarray}
| \phi(\chi, s) \rangle \to \exp[\alpha(\chi, s)] | \phi(\chi, s) \rangle, \\
\langle \varphi(\chi, s) | \to \exp[- \alpha(\chi, s)] \langle \varphi(\chi, s) |,
\end{eqnarray}
the quantity $A(\chi,s)$ transforms as
\begin{eqnarray}
A(\chi,s) \to A(\chi,s) + \frac{\rmd \alpha(\chi, s)}{\rmd s}.
\end{eqnarray}

\begin{figure}
\begin{center}
\includegraphics[width=70mm]{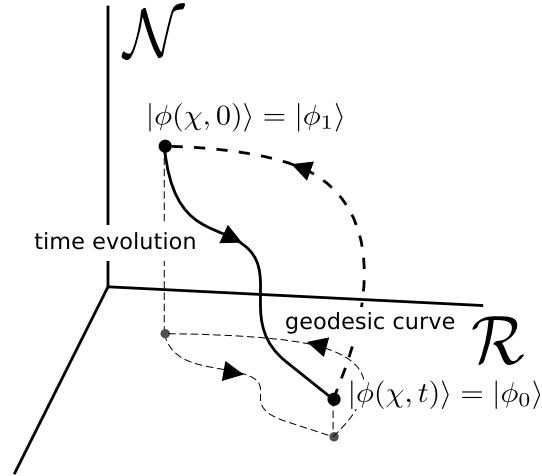}
\end{center}
\caption{A closed curve for the geometric-phase interpretation.
The forward path from $|\phi(\chi,0)\rangle$ to $|\phi(\chi,t) \rangle$
obeys the time evolution of the generating function,
and the backward path from $|\phi_0\rangle = |\phi(\chi,t)\rangle$
to $|\phi_1\rangle = |\phi(\chi,0)\rangle$
is generated by the geodesic curve in $\mathcal{R}$.
}
\label{fig_geometric_interpretation}
\end{figure}

In order to see the geometric characteristics of the phase $\gamma(\chi,t)$
defined by \eref{eq_def_of_gamma},
we consider the `closed' curve in $\mathcal{N}$ as shown in figure~\ref{fig_geometric_interpretation}.
That is, the state develops from $|\phi(\chi,0) \rangle$ to $|\phi(\chi,t) \rangle$
according to the time evolution given by \eref{eq_time_evolution_for_phi},
and the final state $|\phi(\chi,t)\rangle = |\phi_0 \rangle$
is connected to the initial state $|\phi(\chi,0) \rangle = | \phi_1 \rangle$
using a \textit{geodesic} curve.
The construction of the geodesic curve is as follows.
When $|\phi(\chi,s)\rangle$ is a curve in $\mathcal{N}$
and $|u\rangle$ is its tangent vector,
its projection orthogonal to the fiber is 
\begin{eqnarray}
\fl
| u'(\chi,s) \rangle \equiv 
\frac{\rmd}{\rmd s} | \phi(\chi,s) \rangle - A(\chi,s) | \phi(\chi,s) \rangle =
| u(\chi,s) \rangle - | \phi(\chi,s) \rangle
\langle \varphi(\chi,s) | u(\chi,s) \rangle.
\end{eqnarray}
The corresponding projection in the dual space is written as follows:
\begin{eqnarray}
\fl
\langle v'(\chi,s) | \equiv
\frac{\rmd}{\rmd s} \langle \varphi(\chi,s) | + \langle \varphi(\chi,s)| A(\chi,s) =
\langle v(\chi,s) | - \langle \varphi(\chi,s) |
\langle v(\chi,s) | \phi(\chi,s) \rangle.
\end{eqnarray}
Note that $|u\rangle$ and $\langle v|$ are \textit{not} gauge covariant,
but $|u'\rangle$ and $\langle v' |$ transform covariantly under the gauge transformation.
Because $\langle v' | u' \rangle$ is gauge invariant,
we can use it to define a metric on $\mathcal{R}$.
The metric then determines geodesics in $\mathcal{R}$,
which are found by variation of $\int \langle v' | u' \rangle \rmd l$,
where $l$ is an affine parameter determined by the metric.
Hence, the geodesic equation is given as follows:
\begin{eqnarray}
\frac{\rmd}{\rmd s} | u'(\chi,s) \rangle - A(\chi,s) | u'(\chi,s) \rangle = 0, \label{eq_geod} \\
\frac{\rmd}{\rmd s} \langle v'(\chi,s) | + A(\chi,s) \langle v'(\chi,s) | = 0. 
\end{eqnarray}

Using the geodesic curve as defined above,
we can show that the phase $\gamma(\chi,t)$ is written as
\begin{eqnarray}
\gamma(\chi,t) = \oint A(\chi,s) \rmd s.
\label{eq_def_of_gamma_2}
\end{eqnarray}
In order to show \eref{eq_def_of_gamma_2}, let $r(s)$ be a geodesic curve in $\mathcal{R}$ connecting 
$\pi(|\phi_0\rangle)$ to $\pi(|\phi_1\rangle)$, and $0 \leq s \leq 1$.
We define the horizontal lift $|\tilde{\phi}_s\rangle$ 
of this curve, starting from $|\tilde{\phi}_0 \rangle \equiv |\phi_0 \rangle$
and we have $\tilde{A}(\chi,s) = 0$ for arbitrary $s$.
In addition, a gauge transformation 
$|\phi_s \rangle = \exp[\beta_s] |\tilde{\phi}_s\rangle$ is considered,
where $\beta_s$ is chosen so that $\beta_0 = 0$ and 
$\beta_1 = \ln \langle \varphi(\chi,0) | \phi(\chi,t) \rangle$.
Since the geodesic equation \eref{eq_geod} is gauge covariant, 
$|\phi_s\rangle$ is also a geodesic curve,
which connects $|\phi_0\rangle$ and $|\phi_1\rangle$.
Hence, we have 
\begin{eqnarray}
\fl
\int_0^1 A(\chi,s) \rmd s = \int_0^1 \left( \tilde{A}(\chi,s) + \frac{\rmd \beta_s}{\rmd s} \right) \rmd s
= \int_0^1 \frac{\rmd \beta_s}{\rmd s} \rmd s = \ln \langle \varphi(\chi,0) | \phi(\chi,t) \rangle.
\end{eqnarray}
Since the quantity $A(\chi,s)$ is always zero
during the time evolution from $|\phi(\chi,0)\rangle$ to $|\phi(\chi,t)\rangle$,
we can conclude that the integral in \eref{eq_def_of_gamma_2}
on the closed line defined in figure~\ref{fig_geometric_interpretation} 
adequately gives the phase $\gamma$ in \eref{eq_def_of_gamma}.

\section{Concluding remarks}

In the present paper, we gave a general discussion of a noncyclic geometric phase for counting statistics,
which enables us to extend the adiabatic discussions in 
\cite{Sinitsyn2008a,Sinitsyn2010} to nonadiabatic cases.
Firstly, we clarified that the counting statistics
can be reformulated so as to compare a bra state at the initial time and a ket state at an arbitrary time.
Secondly, removing the dynamical contribution from the cumulant generating function,
a remaining phase was introduced.
The dynamical and remaining phases would correspond, respectively, to cumulant generating functions
with different meanings.
Thirdly, it was clarified that the remaining phase is also mathematically meaningful;
the remaining phase can be interpreted as a `geometric phase'.
In order to interpret the remaining phase as the geometric one,
a geodesic curve was used,
and we obtain a closed contour of the integral;
the contour of the integral is given by the actual evolution of the generating function
and a geodesic curve which connects the initial and final states of the generating function backwardly.
In addition, the remaining phase is clearly gauge invariant,
and it depends only on the geometric path in the base space;
it does not depend on its rate of traversal.
Hence, the definition of the remaining phase readily implies a geometric interpretation.

We here note that 
the discussions using the geodesic curves are used only for the justification
of the geometric property of the phase.
In practice, there is no need to estimate the geometric phase by finding the geodesic curves;
it is enough to use \eref{eq_def_of_gamma}.
Hence, in order to obtain the dynamical phase and the geometric phase,
only the solutions of the generating functions are necessary.

In section 1, we noted that
a mathematically interesting structure may also be physically-meaningful.
For cyclic cases, as denoted in section 3,
the mathematical structure is indeed physically-meaningful;
the geometric phase gives information about the pumping phenomena induced by periodic perturbations
\cite{Sinitsyn2007,Sinitsyn2007a,Sinitsyn2008,Sinitsyn2009,Ohkubo2008,Ohkubo2008a}.
It is still unclear whether the geometric phase is also important
for general nonequilibrium states.
In order to confirm it, studies for specific models would be useful and needed;
these are out of the scope of the present paper, and will be clarified in future works.
In what follows, we will give some discussions
to apply the geometric phase to studies of nonequilibrium states.

When adiabatic cases are considered,
the dynamical phase gives simply
information about the number of specific transitions in quasi-static cases.
Hence, the noncyclic geometric phase will correspond to
an additional contribution from the time-dependent transition matrix.
Actually, in \cite{Sinitsyn2008a,Sinitsyn2010},
such effects in adiabatic cases are discussed for a simple two-state model.
This means that even in a quasi-static evolution,
the total number of transitions
is not given by a quasi-static contribution.
It would sometimes be beneficial to discuss these two contributions separately,
and the noncyclic geometric phase discussions divide them naturally even in nonadiabatic cases.

Next, let us consider transitions between two nonequilibrium steady states.
In such cases, it has been pointed out that
a concept of `excess' quantities would be important to construct useful identities or relations;
excess heat or entropy production have already been discussed in a few contexts
\cite{Hatano2001,Oono1998,Speck2005,Esposito2007,Komatsu2008,Komatsu2008a,Komatsu2009,Esposito2010}.
Here, the `excess' quantity means an additional quantity caused by time-dependent
perturbations or external fields,
and it gives no contribution if the state is a steady one.
For example, an excess heat has been explicitly introduced and calculated in a Langevin system,
and the concept is deeply related to the Hatano-Sasa relation \cite{Hatano2001},
which can be considered as an extension of the Jarzynski equality to nonequilibrium cases.
The excess heat is estimated by subtracting a steady heat, the so-called housekeeping heat,
from the total heat.
The housekeeping heat is needed to maintain the steady state out of equilibrium.
One of the remarkable characteristics of the excess heat
is that it is a finite quantity.
In contrast, the total and the housekeeping heat during time $t$ diverge when $t \to \infty$.
It may be plausible to use such finite quantities 
in order to discuss nonequilibrium steady states \cite{Oono1998}.
%---- I modified the following sentences
An important point here is that 
a total quantity is divided into two parts,
i.e., the housekeeping one and the excess one.
In the noncyclic geometric phase discussions,
we observed that the total phase is naturally divided into two parts,
i.e., the dynamical one and the geometric one, as described above.
In this sense, one may expect that 
this natural division of the total phase have
some physical meanings, as in the case of the excess quantity.
Although further studies will be needed to clarify the physical meanings
of the noncyclic geometric phase,
we hope that such discussions based on the noncyclic geometric phase
would give deep insights for nonequilibrium states.

\section*{Acknowledgments}
This work was supported in part by grant-in-aid for scientific research 
(Grants No.~20115009 and No.~21740283)
from the Ministry of Education, Culture, Sports, Science and Technology (MEXT), Japan.
T.E. is supported by a Government Scholarship from the MEXT.

\appendix

\section{Generating function for counting statistics}

Although counting statistics has been discussed in various contexts
(for example in \cite{Gopich2005}),
we here give a rederivation of the generating function 
used in counting statistics for the reader's convenience.

In the framework of counting statistics, the quantity of interest is the number of target transitions.
The target transitions can be chosen arbitrarily.
For example, if we consider a difference between the number of transitions
from state $i$ to state $j$ (we denote this transition as $i \to j$)
and that of transition $j \to i$
we can calculate a net `current' of $i \to j$.
The information about such `current' plays an important role in discussing nonequilibrium states.

Let $\{K_{nm}(t)\}$ be a time-dependent transition matrix.
Although one can calculate the statistics for multiple target transitions in general,
we here derive the generating function for counting the number of events 
of a {\it specific} target transition $i_\mathrm{A} \to j_\mathrm{A}$, for simplicity.
The generalization to multiple transitions is straightforward \cite{Gopich2005}.

Let us denote the probability, with which the system starts from state $m$ and finishes in state $n$
with $N_\mathrm{A}$ being the number of target transitions $i_\mathrm{A} \to j_\mathrm{A}$ during time $t$, 
as $P_{nm}(N_\mathrm{A}|t)$.
In order to calculate the probability $P_{nm}(N_\mathrm{A}|t)$, we here define
a probability $G_{kl}'(t)$ with which the system evolves from state $l$ to state $k$, 
provided no $i_\mathrm{A} \to j_\mathrm{A}$ transitions occur during time $t$.
By using the probability $G_{kl}'(t)$,
the probability $P_{nm}(N_\mathrm{A}|t)$ is calculated as
\begin{eqnarray}
\fl
P_{nm}(N_\mathrm{A} | t ) = G_{n j_\mathrm{A}}'(t) \ast 
\underbrace{K_{j_\mathrm{A} i_\mathrm{A}}(t) G_{i_\mathrm{A} j_\mathrm{A}}'(t) \ast \cdots 
\ast K_{j_\mathrm{A} i_\mathrm{A}}(t) G_{i_\mathrm{A} j_\mathrm{A}}' (t)
}_{N_\mathrm{A} - 1}
\ast K_{j_\mathrm{A} i_\mathrm{A}}(t) G_{i_\mathrm{A} m}' (t),
\end{eqnarray}
where $g_1(t) \ast g_2(t) \equiv \int_0^t g_1(t-t') g_2(t') \rmd t'$ denotes the convolution.
This formulation means that an occurrence of the target transition $i_\mathrm{A} \to j_\mathrm{A}$
is sandwiched in between situations with no occurrence of the target transition,
and it is repeated $N_\mathrm{A}$ times.

Next, we construct the generating function $\tilde{f}_{nm}(\chi,t)$
of the probability $P_{nm}(N_\mathrm{A}|t)$:
\begin{eqnarray}
\tilde{f}_{nm}(\chi,t) = \sum_{N_\mathrm{A}=0}^\infty \rme^{\chi N_\mathrm{A}} P_{nm}(N_\mathrm{A}|t).
\end{eqnarray}
That is, the generating function $\tilde{f}_{nm}(\chi,t)$ 
gives the statistics of the number of transition $i_\mathrm{A} \to j_\mathrm{A}$ during time $t$
under the condition that the system starts from state $m$ and ends in state $n$.
The generating function $\tilde{f}_{nm}(\chi,t)$ 
satisfies the following integral equation
\begin{eqnarray}
\tilde{f}_{nm}(\chi,t) = G_{nm}'(t) +
\int_0^t G_{n j_\mathrm{A}}' (t-t') \rme^{\chi} K_{j_\mathrm{A} i_\mathrm{A}}(t')
\tilde{f}_{i_\mathrm{A} m} (\chi,t') \rmd t',
\end{eqnarray}
and obeys the following time-evolution equation
\begin{eqnarray}
\fl
\frac{\rmd}{\rmd t} \tilde{f}_{nm} (\chi,t) 
&= \sum_i K_{ni}(t) G_{im}'(t) 
- \delta_{n, j_\mathrm{A}} K_{j_\mathrm{A} i_\mathrm{A}}(t) G_{i_\mathrm{A} m}'(t)
+ \rme^{\chi} G_{n j_\mathrm{A}}'(0) K_{j_\mathrm{A} i_\mathrm{A}}(t)  \tilde{f}_{i_\mathrm{A} m}(t)
\nonumber \\
\fl
& \quad + \int_0^t \left( \frac{\rmd}{\rmd t} G_{n j_\mathrm{A}}'(t-t')\right)
\rme^{\chi} K_{j_\mathrm{A} i_\mathrm{A}}(t') \tilde{f}_{i_\mathrm{A} m}(t') dt' \nonumber \\
\fl
&=
\sum_i K_{ni}(t) \tilde{f}_{im}(\chi,t) - \delta_{n, j_\mathrm{A}} (1-\rme^{\chi}) 
K_{j_\mathrm{A} i_\mathrm{A}}(t) \tilde{f}_{i_\mathrm{A} m}(\chi,t),
\label{eq_appendix_time_evolution_for_fnm}
\end{eqnarray}
where $\tilde{f}_{nm}(\chi,0) = \delta_{n,m}$.
In order to show \eref{eq_appendix_time_evolution_for_fnm},
we used the following two facts:
Firstly, 
the probability of no target transitions, $G_{nm}'(t)$, obeys
\begin{eqnarray}
\frac{\partial}{\partial t} G_{nm}'(t) = \sum_i K_{ni} (t) G_{im}'(t)
- \delta_{n, j_\mathrm{A}} K_{j_\mathrm{A} i_\mathrm{A}} (t) G_{i_\mathrm{A} m}'(t),
\end{eqnarray}
where $G_{nm}'(0) = \delta_{n,m}$.
Secondly, 
the derivative of the convolution is given by 
\begin{eqnarray}
\frac{\partial}{\partial t} \int_0^t g_1(t-t') g_2(t') \rmd t'
= g_1(0) g_2(t) + \int_0^t  \left( \frac{\partial}{\partial t} g_1(t - t') \right) g_2(t') \rmd t'.
\end{eqnarray}

Using the generating function $\tilde{f}_{nm}(\chi,t)$,
we construct the restricted generating function $f_n(\chi,t)$ in 
\eref{eq_redef_of_total_generating_function} and  \eref{eq_gf_original}
as follows:
\begin{eqnarray}
f_n(\chi,t) = \sum_m \tilde{f}_{nm}(\chi,t) p_m(0),
\label{appendix_eq_def_fn}
\end{eqnarray}
where $p_m(0)$ is a probability distribution at initial time $t=0$.
Hence the generating function $F(\chi,t)$ is given by \eref{eq_redef_of_total_generating_function}.
In addition, from \eref{eq_appendix_time_evolution_for_fnm} and \eref{appendix_eq_def_fn},
 the restricted generating function $f_n(\chi,t)$ satisfies
\begin{eqnarray}
\fl
\frac{\rmd}{\rmd t} f_{n}(\chi,t)
= \sum_{i} K_{ni}(t) f_{i}(\chi,t) - \delta_{n,j_\mathrm{A}} (1-\rme^{\chi}) 
K_{j_\mathrm{A}i_\mathrm{A}}(t)
f_{i_\mathrm{A}}(\chi,t),
\end{eqnarray}
and these equations should be solved with initial conditions 
$f_n(\chi,0) = \sum_m \tilde{f}_{nm}(\chi,0) p_m(0)  = p_n(0)$.


\section*{References}
\begin{thebibliography}{99}



% GC type and integral fluctuation theorem
\bibitem{Gallavotti1995}  Gallavotti G and Cohen E G D 1995 {\it Phys. Rev. Lett.} {\bf 74} 2694
\bibitem{Gallavotti1995a} Gallavotti G and Cohen E G D 1995 {\it J. Stat. Phys.} {\bf 80} 931
\bibitem{Kurchan1998} Kurchan J 1998 {\it J. Phys. A: Math. Gen.} {\bf 31} 3719
\bibitem{Crooks1998} Crooks G E 1998 {\it J. Stat. Phys.} {\bf 90} 1481
\bibitem{Crooks1999} Crooks G E 1999 {\it Phys. Rev. E} {\bf 60} 2721
\bibitem{Lebowitz1999} Lebowitz J L and Spohn H 1999 {\it J. Stat. Phys.} {\bf 95} 333
\bibitem{Maes1999} Maes C 1999 {\it J. Stat. Phys.} {\bf 95} 367
\bibitem{Crooks2000} Crooks G E 2000 {\it Phys. Rev. E} {\bf 61} 2361
\bibitem{Gaspard2004} Gaspard P 2004 {\it J. Chem. Phys.} {\bf 120} 8898
\bibitem{Seifert2005} Seifert U 2005 {\it Phys. Rev. Lett.} {\bf 95} 040602

% Jarzynski equality
\bibitem{Jarzynski1997} Jarzynski C 1997 {\it Phys. Rev. Lett.} {\bf 78} 2690
\bibitem{Jarzynski1997a} Jarzynski C 1997 {\it Phys. Rev. E} {\bf 56} 5018

% Hatano-Sasa
\bibitem{Hatano2001} Hatano T and Sasa S 2001 {\it Phys. Rev. Lett.} {\bf 86} 3463

% generalized fluctuation theorem
\bibitem{Esposito2007} Esposito M, Harbola U and Mukamel S 2007 {\it Phys. Rev. E} {\bf 76} 031132
\bibitem{Seifert2008} Seifert U 2008 {\it Eur. Phys. J. B} {\bf 64} 423
\bibitem{Ohkubo2009} Ohkubo J 2009 {\it J. Phys. Soc. Jpn.} 123001




% counting statistics in classical stochastic process
\bibitem{Bicout1999} Bicout D J and Rubin R J 1999 {\it Phys. Rev. E} {\bf 59} 913
\bibitem{Gopich2003} Gopich I V and Szabo A 2003 {\it J. Chem. Phys.} {\bf 118} 454
\bibitem{Gopich2005} Gopich I and Szabo A 2005 {\it J. Chem. Phys.} {\bf 122} 014707
\bibitem{Sung2005} Sung J and Silbey R J 2005 {\it Chem. Phys. Lett.} {\bf 415} 10
\bibitem{Gopich2006} Gopich I V and Szabo A 2006 {\it J. Chem. Phys.} {\bf 124} 154712

% geometric phase for counting statistics (cyclic)
\bibitem{Sinitsyn2007} Sinitsyn N A and Nemenman I 2007 {\it Europhys. Lett.} {\bf 77} 58001
\bibitem{Sinitsyn2007a} Sinitsyn N A and Nemenman I 2007 {\it Phys. Rev. Lett.} {\bf 99} 220408
% geometric phase for counting statistics (others: holonomy, review)
\bibitem{Sinitsyn2008} Sinitsyn N A and Saxena A 2008 {\it J. Phys. A: Math. Theor.} {\bf 41} 392002
\bibitem{Sinitsyn2009} Sinitsyn N A 2009 {\it J. Phys. A: Math. Theor.} {\bf 42} 193001
% geometric phase ... (nonadiabatic)
\bibitem{Ohkubo2008} Ohkubo J 2008 {\it J. Stat. Mech.} P02011
\bibitem{Ohkubo2008a} Ohkubo J 2008 {\it J. Chem. Phys.} {\bf 129} 205102


% noncyclic geometric phase
\bibitem{Samuel1988} Samuel J and Bhandari R 1988 {\it Phys. Rev. Lett.} {\bf 60} 2339
% check!
\bibitem{Aitchison1992} Aitchison I J R and Wanelik K 1992 {\it Proc. R. Soc. A} {\bf 439} 25
% check!
\bibitem{Mukanda1993} Mukunda N and Simon R 1993 {\it Ann. Phys.} {\bf 228} 205
\bibitem{Mostafazadeh1999} Mostafazadeh A 1999 {\it J. Phys. A: Math. Gen.} {\bf 32} 8157


% geometric phase for counting statistics (noncyclic)
\bibitem{Sinitsyn2008a} Sinitsyn N A 2008 {\it LANL technical report} LA-UR-08-04425
\bibitem{Sinitsyn2010} Sinitsyn N A and Nemenman I 2010 {\it Preprint} arXiv:1001.4212


\bibitem{Weiss_book} Weiss U 2008 {\it Quantum Dissipative Systems, 3rd edition} 
(Singapore : World Scientific)
% note
\bibitem{note1} It is also possible to obtain more detailed statistics.
For example, when a transition $2 \to 1$ can be divided into two parts,
i.e., $K_{12}(t) = k_{12}^{(1)}(t) + k_{12}^{(2)}(t)$,
one can calculate the statistics of only the transition $k_{12}^{(2)}(t)$.
In this case, the corresponding element of the modified transition matrix is
$k_{12}^{(1)}(t) + k_{12}^{(2)}(t) \rme^{\chi}$.
As for these generalizations, e.g., see \cite{Gopich2005}.


% complex geometric phase
\bibitem{Garrison1988} Garrison J C and Wright E M 1988 {\it Phys. Lett. A} {\bf 128} 177

% excess flow
\bibitem{Oono1998} Oono Y and Paniconi M 1998 {\it Prog. Theor. Phys. Supple.} {\bf 130} 29
\bibitem{Speck2005} Speck T and Seifert U 2005 {\it J. Phys. A: Math. Gen.} {\bf 38} L581
\bibitem{Komatsu2008} Komatsu T S and Nakagawa N 2008 {\it Phys. Rev. Lett.} {\bf 100} 030601
\bibitem{Komatsu2008a} Komatsu T S, Nakagawa N, Sasa S I and Tasaki H 2008
{\it Phys. Rev. Lett.} {\bf 100} 230602
\bibitem{Komatsu2009} Komatsu T S, Nakagawa N, Sasa S I and Tasaki H 2009 
{\it J. Stat. Phys.} {\bf 134} 401
\bibitem{Esposito2010} Esposito M and Broeck C V den 2010 {\it Phys. Rev. Lett.} {\bf 104} 090601



\end{thebibliography}
\end{document}